\documentclass[12pt]{article}\usepackage{graphicx}
\begin{document}
%\titlerunning{Title running}
\begin{center}
{\Large\bf \boldmath Light-cone expansion of heavy-to-light form factors}
%\end{center}

\vspace*{6mm}
%\begin{center}
{Wolfgang Lucha$^a$, Dmitri Melikhov$^{a,b}$, and Silvano
Simula$^c$}\\{\small \it $^a$ HEPHY, Austrian Academy of Sciences,
Nikolsdorfergasse 18, A-1050, Vienna, Austria}\\{\small \it $^b$
SINP, Moscow State University, 119991, Moscow, Russia}\\{\small\it
$^c$ INFN, Sezione di Roma III, Via della Vasca Navale 84,
I-00146, Roma, Italy}
\end{center}

\vspace*{3mm}

\begin{abstract}
We present the results of our recent systematic study of the
light-cone expansion~of heavy-to-light transition form factors in
a model with scalar constituents \cite{lms_lcsr}. We~show that the
higher-twist contributions (represented in this model by
off-light-cone effects) all have the same behaviour in the $1/m_Q$
expansion. The suppression parameter of~the higher-twist
contributions compared to the lower-twist contributions is, in
general,~the inverse Borel parameter $\beta$. The only exception
here is the case of the leading and the subleading twists: they
are of the same order in $1/\beta$ because of an extra
suppression~of the leading-twist contribution to the form factor.
\end{abstract}
\vspace*{6mm}

Light-cone (LC) sum rules \cite{lcsr} belong to the most widely
used approaches for calculating hadron form factors in QCD. The
form factor of an individual bound state obtained from~a LC sum
rule depends on two ingredients: (i) the field-theoretic
calculation of the relevant correlator by constructing its LC
expansion in terms of hadron distribution amplitudes (DA) of
increasing twist, and (ii) the technical ``extraction procedure''
(cutting the correlator and determining the effective continuum
threshold), which introduces a systematic error~into~the extracted
form factor (for a recent study of the serious issue of the
systematic errors, see~\cite{lms_sr}).

In QCD one can calculate only a few terms of the LC expansion of
the correlator; it is impossible to study the full series and to
estimate the typical size of the omitted higher-twist effects.
Therefore, we have systematically analysed the LC expansion in a
model~with~scalar constituents: a heavy ``quark'' field $Q$ (of
mass $m_Q$) and a light ``quark'' field $\varphi$ (of mass~$m$)
interacting by exchange of a massless boson in ladder
approximation \cite{lms_lcsr}. The basic object here is the
heavy-to-light correlator
\begin{eqnarray}
\Gamma(p^2,q^2)&=&i \int d^4x \exp({ipx})\langle M(p')|T
\left(\varphi(x)Q(x)Q(0)\varphi(0)\right)|0\rangle.
\end{eqnarray}
One should (i) obtain the dispersion representation in $p^2,$
\begin{eqnarray}
\Gamma(p^2,q^2)=\int \frac{ds}{s-p^2-i0}\Delta(s,q^2),
\end{eqnarray}
and (ii) perform the Borel transform $1/(s-p^2)\to
\exp\left[-s/(2m_Q\beta)\right]$ ($\beta\ll m_Q$) and relate the
{\it cut} Borel image to the form factor of interest:
\begin{eqnarray}
f_{M_Q}\,F_{M_Q\to
M}(q^2)=\exp\left(\frac{M_Q^2}{2m_Q\beta}\right)\Gamma(\beta,q^2,s_0)\equiv
\int\limits_{(m_Q+m)^2}^{s_0}
{ds}\exp\left(-\frac{s-M_Q^2}{2m_Q\beta}\right)\Delta(s,q^2).
\end{eqnarray}
Here, $s_0=(m_Q+z_{\rm eff})^2$ is the {\rm effective} continuum
threshold to be fixed by some criterion. For our analysis it is
essential that $z_{\rm eff}$ is a constant which remains finite in
the limit $m_Q\to\infty$.

\newpage

Upon performing the factorization, the spectral density
$\Delta(s,q^2)$ may be found via the~LC expansion of the soft
Bethe--Salpeter (BS) amplitude of the light-quark bound state
$M(p')$. The LC expansion of the soft BS amplitude at the
factorization scale $\lambda$ may be written~in~the
form\footnote{In the ladder-approximation model with scalar
constituents, one may obtain the correlator $\Gamma$ via the~BS
amplitude without performing the factorization; also the LC
contribution to the correlator $\Gamma_0$~can~be found. However,
any proper treatment of the full LC expansion of the correlator
still requires factorization.}
\begin{eqnarray}
\label{4} \Psi_{\rm soft}(x,p'|\lambda)\equiv \langle M(p') |T
\varphi(x)\varphi(0)|0\rangle_{\lambda}=
\sum_{n=0}^{\infty}(x^2)^n\int\limits_0^1 d\xi \exp({-i
p'x\xi})\phi_n(\xi,\lambda).
\end{eqnarray}
Here $\phi_n(\xi,\lambda)=C_n(\lambda)(m^2)^n\xi(1-\xi),$
$n=0,1,2,\dots,$ are the DAs of increasing twist~involving
calculable functions $C_n(\lambda)$ of the factorization scale
$\lambda.$ The end-point behaviour of the DAs
$\phi_n(\xi,\lambda)$ depends on the inter-``quark'' interaction.

The LC expansion of the BS amplitude (\ref{4}) generates the LC
expansion of the correlator. Below, we give the corresponding
expressions for $q^2=0.$ For the uncut correlator
($s_0\to\infty$), which is {\em not\/} related to the contribution
of the individual bound state and accordingly {\em not\/}~to the
form factor of interest, neglecting terms of order ${\rm O}(M^2)$
one obtains
\begin{eqnarray}
\Gamma(p^2|\lambda)&=& \frac1{(2\pi)^4}\int d^4k\, d^4x
\,\frac{\exp[i(p-k)x]}{m_Q^2-k^2-i0}
\sum_{n=0}^{\infty}(x^2)^n\int\limits_0^1 d\xi \exp({-i
p'x\xi})\phi_n(\xi,\lambda) \nonumber \\ &=& \int\limits_0^1
\frac{d\xi \,\phi_0(\xi,\lambda)}{m_Q^2-p^2(1-\xi)} -8m_Q^2
\int\limits_0^1\frac{d\xi \,\phi_1(\xi,\lambda)}
{\left[m_Q^2-p^2(1-\xi)\right]^3}+\cdots\equiv
\Gamma_0+\Gamma_1+\cdots.\qquad
\end{eqnarray}
The Borelized uncut correlator takes the following form:
$$\Gamma(\beta|\lambda)\simeq \exp\left(-{m_Q}/{2\beta}\right)
\int\limits_0^1 \frac{d\xi}{1-\xi}
\left[\phi_0(\xi,\lambda)-\frac{\phi_1(\xi,\lambda)}{\beta^2(1-\xi)^2}
+\cdots \right] \exp\left(-\frac{m_Q \xi}{2\beta(1-\xi)}\right).$$
Obviously, the main contribution to the integral arises from the
end-point region $\xi\simeq \beta/m_Q$. For this uncut correlator
the behaviour of the contributions of increasing twist is found
to~be
\begin{eqnarray*}
&&\Gamma_0(\beta)\sim
\exp\left(-\frac{m_Q}{2\beta}\right)\frac{\beta^2}{m_Q^2},\qquad
\Gamma_1(\beta)\sim
\exp\left(-\frac{m_Q}{2\beta}\right)\frac{m^2}{m_Q^2},\\
&&\Gamma_{n+1}(\beta)\sim
\Gamma_n(\beta)\left(\frac{m^2}{\beta^2}\right)^n,\qquad
n=0,1,2,\ldots.
\end{eqnarray*}
Consequently, in the {\em uncut\/} correlator contributions of
higher twist are suppressed by powers of $m^2/\beta^2$ compared to
those of lower twist.

For the cut correlator, which {\em is\/} related to the form
factor of interest, one should take~care when applying the cut in
the dispersion representation, which leads to surface terms
\cite{lms_lcsr}. The surface terms modify the leading behaviour of
$\Gamma_0$ while leaving the leading behaviour~of~the higher-twist
contributions unchanged:
\begin{eqnarray*}
&&\Gamma_0(\beta,z_{\rm eff})\sim
\exp\left(-\frac{m_Q}{2\beta}\right)\frac{z^2_{\rm
eff}}{m_Q^2}\nonumber, \qquad \Gamma_1(\beta,z_{\rm eff})\sim
\exp\left(-\frac{m_Q}{2\beta}\right)\frac{m^2}{m_Q^2}, \qquad\\
&&\Gamma_{n+1}(\beta,z_{\rm eff})\sim \Gamma_n(\beta,z_{\rm
eff})\left(\frac{m^2}{\beta^2}\right)^n,\qquad n=1,2,3,\ldots.
\end{eqnarray*}
Note that the parameter $z_{\rm eff}$ here is a fixed parameter
(in our model $\sim m$, in QCD $\sim \Lambda_{\rm QCD}$) dictated
by dynamics. Thus, the leading and the subleading twists are of
the same order~of magnitude. For higher twists, the twist
hierarchy is preserved in the cut correlator~related~to the form
factor, similar to the case of the uncut correlator.

In \cite{lms_lcsr} we have calculated, without invoking the LC
expansion, the full correlator by~use~of the known solution for
the BS amplitude in ladder approximation and, separately, the LC
contribution to the correlator. We found that numerically the
leading twist provides about 70\% of the full correlator. The
contributions of the higher twists (mainly, of the subleading
twist), however, stay at the order of 30\% for all values of the
parameter $\beta$ and of the~mass~$m_Q$ of the heavy ``quark.''

In summary, in the LC sum-rule approach to heavy-to-light form
factors near $q^2=0$ the contributions of all twists exhibit the
same behaviour in the $1/m_Q$ expansion. Higher-twist
contributions are suppressed by powers of the parameter
$\Lambda_{\rm QCD}/\beta$, with the exception of the leading- and
the subleading-twist contributions, which exhibit the same
behaviour in $1/\beta$.

\noindent However, since the ground-state contribution is enhanced
for small $\beta$, one would like~to~know the correlator for the
smallest possible values of $\beta$, where the inclusion of
higher-twist effects is mandatory. The off-LC and other
higher-twist effects in weak decays of heavy mesons in QCD deserve
a detailed investigation: for the LC sum-rule method the
corresponding~DAs are just external objects, which should be
provided by other nonperturbative approaches.~In particular, the
combination of LC sum rules with techniques based on bound-state
equations and the constituent quark picture \cite{melikhov} (which
successfully describe heavy-meson decays)~may prove to be
promising.

\vspace{3ex}\noindent {\bf Acknowledgements.} We thank the
organizers for the invitation to this highly interesting meeting.
D.~M.\ was supported by FWF project P20573 and by RFBR project
07-02-00551.


\begin{thebibliography}{99}\itemsep -1mm
\bibitem{lms_lcsr}W.~Lucha, D.~Melikhov, and S.~Simula,
Phys.~Rev.~D {\bf 75}, 096002 (2007); Phys.~Atom.\ Nucl.~{\bf 71},
545 (2008).
\bibitem{lcsr}I.~I.~Balitsky, V.~M.~Braun, and A.~V.~Kolesnichenko,
Nucl.~Phys.~B {\bf 312}, 509 (1989); V.~M.~Braun and I.~Filyanov,
Z.~Phys.~C {\bf 44}, 157 (1989); V.~I.\ Chernyak and
I.~R.~Zhitnitsky, Nucl.~Phys.~B {\bf 345}, 137 (1990).
\bibitem{lms_sr}W.~Lucha, D.~Melikhov, and S.~Simula,
Phys.~Rev.~D {\bf 76}, 036002 (2007); Phys.~Lett.~B {\bf 657}, 148
(2007); Phys.~Atom.~Nucl.~{\bf 71}, 1461 (2008); W.~Lucha and
D.~Melikhov,~Phys.\ Rev.~D {\bf 73}, 054009 (2006);
Phys.~Atom.~Nucl.~{\bf 70}, 891 (2007).
\bibitem{melikhov}V.~Anisovich et al., Nucl. Phys.~A {\bf 544}, 747
(1992); D.~Melikhov, Phys.~Rev.~D {\bf 53},~2460 (1996);
Eur.~Phys.~J.~direct C {\bf 4}, 2 (2002) [hep-ph/0110087];
D.~Melikhov and B.~Stech, Phys.~Rev.~D {\bf 62}, 014006 (2000);
D.~Melikhov and S.~Simula, Eur.~Phys.~J.~C {\bf 37}, 437 (2004).
\end{thebibliography}
\end{document}